\documentclass[11pt]{article}

\usepackage[margin=1in]{geometry}
\usepackage[utf8]{inputenc}
\usepackage[T1]{fontenc}
\usepackage{times}
\usepackage{booktabs}
\usepackage{graphicx}
\usepackage{amsmath}
\usepackage{amssymb}
\usepackage{xspace}
\usepackage{url}
\usepackage{natbib}
\usepackage[hidelinks]{hyperref}

\newcommand{\ArchiveRows}{6,638,350\xspace}
\newcommand{\ArchiveEntities}{104,459\xspace}
\newcommand{\ArchiveSelfParent}{89,962\xspace}
\newcommand{\ArchiveGenuineLinks}{14,288\xspace}
\newcommand{\ArchiveConflicting}{1,005\xspace}
\newcommand{\ArchiveSelfParentPct}{86.3\%\xspace}
\newcommand{\AuditRecords}{4,806\xspace}
\newcommand{\AuditChildRecords}{4,424\xspace}
\newcommand{\AuditSelfParentPct}{82.8\%\xspace}
\newcommand{\AuditGenuineLinks}{760\xspace}
\newcommand{\AuditMultiParent}{1,065\xspace}
\newcommand{\AuditMaxParents}{9\xspace}

\newcommand{\ReconBoth}{248\xspace}
\newcommand{\ReconAgreeUei}{64.9\%\xspace}
\newcommand{\ReconAgreeName}{73.8\%\xspace}
\newcommand{\ReconAgreeToken}{81.5\%\xspace}
\newcommand{\ReconAgreeNameCI}{[68.0, 78.9]\xspace}
\newcommand{\ReconDisagree}{46\xspace}
\newcommand{\ReconDisagreePct}{18.5\%\xspace}
\newcommand{\ReconOnlyOne}{264\xspace}
\newcommand{\NvIdenticalRaw}{4,850\xspace}
\newcommand{\NvInformative}{9,438\xspace}
\newcommand{\NvInvisiblePctInformative}{38.3\%\xspace}
\newcommand{\BenchFamilies}{10,307\xspace}
\newcommand{\BenchPairs}{54,864\xspace}
\newcommand{\BenchPositives}{13,716\xspace}
\newcommand{\BenchNegatives}{41,148\xspace}
\newcommand{\BenchDroppedArtifact}{458\xspace}
\newcommand{\BenchDroppedSovereign}{114\xspace}
\newcommand{\BenchExcludedParents}{8\xspace}
\newcommand{\BenchInspected}{28\xspace}
\newcommand{\BenchRetained}{20\xspace}
\newcommand{\BenchSovereignAll}{74\xspace}
\newcommand{\BenchDropInvisiblePct}{84.7\%\xspace}
\newcommand{\BenchSeed}{20260827\xspace}
\newcommand{\BenchPosVisible}{2,953\xspace}
\newcommand{\BenchPosVisiblePct}{21.5\%\xspace}
\newcommand{\BenchPosIdentical}{7,541\xspace}
\newcommand{\BenchPosIdenticalPct}{55.0\%\xspace}
\newcommand{\BenchPosInvisible}{3,222\xspace}
\newcommand{\BenchPosInvisiblePct}{23.5\%\xspace}
\newcommand{\BenchPosTrain}{8,229\xspace}
\newcommand{\BenchPosTest}{2,743\xspace}
\newcommand{\BenchPosVal}{2,744\xspace}
\newcommand{\BenchPairsTrain}{32,916\xspace}
\newcommand{\BenchPairsTest}{10,972\xspace}
\newcommand{\BenchPairsVal}{10,976\xspace}
\newcommand{\BaseIdenticalRate}{97.3\%\xspace}
\newcommand{\BaseIdenticalF}{98.6\xspace}
\newcommand{\BaseVisibleRate}{10.2\%\xspace}
\newcommand{\BaseVisibleF}{18.5\xspace}
\newcommand{\BaseInvisibleRate}{16.9\%\xspace}
\newcommand{\BaseInvisibleF}{29.0\xspace}
\newcommand{\BestRecallIdentical}{100.0\%\xspace}
\newcommand{\BestRecallIdenticalCI}{[99.8, 100.0]\xspace}
\newcommand{\BestRecallVisible}{49.9\%\xspace}
\newcommand{\BestRecallVisibleCI}{[45.8, 54.0]\xspace}
\newcommand{\BestRecallInvisible}{4.2\%\xspace}
\newcommand{\BestRecallInvisibleCI}{[2.9, 6.1]\xspace}
\newcommand{\AnyInvisibleRecallMax}{4.7\%\xspace}
\newcommand{\AnyInvisibleRecallMin}{0.0\%\xspace}
\newcommand{\BestAP}{78.7\xspace}
\newcommand{\BestMethod}{TF--IDF char 3-gram\xspace}
\newcommand{\BestOverallF}{76.8\xspace}
\newcommand{\BestIdenticalF}{98.7\xspace}
\newcommand{\BestVisibleF}{56.3\xspace}
\newcommand{\BestInvisibleF}{8.0\xspace}
\newcommand{\BestSpendRecall}{60.1\%\xspace}
\newcommand{\BestOverallRecall}{67.3\xspace}
\newcommand{\BestOverallPrec}{89.6\xspace}
\newcommand{\AnyInvisibleFMax}{9.0\xspace}
\newcommand{\AnyInvisibleFMin}{0.0\xspace}
\newcommand{\BestInvisibleMethod}{MiniLM embedding\xspace}
\newcommand{\VisibilityGap}{89.7\xspace}
\newcommand{\NumMethods}{5\xspace}
\newcommand{\NumStringMethods}{4\xspace}
\newcommand{\EmbModel}{MiniLM-L6-v2\xspace}
\newcommand{\EmbNames}{18,146\xspace}
\newcommand{\EmbDim}{384\xspace}
\newcommand{\EmbOverallF}{78.5\xspace}
\newcommand{\EmbIdenticalF}{98.7\xspace}
\newcommand{\EmbVisibleF}{64.3\xspace}
\newcommand{\EmbInvisibleF}{9.0\xspace}
\newcommand{\BestStringInvisibleF}{8.0\xspace}
\newcommand{\EmbOverStringInvisible}{1.0\xspace}
\newcommand{\BestFprHardString}{4.6\%\xspace}
\newcommand{\BestFprRandom}{0.0\%\xspace}
\newcommand{\BestFprSameBlock}{3.2\%\xspace}
\newcommand{\BlkEntities}{114,230\xspace}
\newcommand{\BlkPairSpace}{6,524,189,335\xspace}
\newcommand{\BlkTokenPC}{68.7\%\xspace}
\newcommand{\BlkTokenRR}{99.98\%\xspace}
\newcommand{\BlkTokenInvisible}{0.00\%\xspace}
\newcommand{\BlkUnionPC}{76.2\%\xspace}
\newcommand{\BlkUnionInvisible}{6.8\%\xspace}
\newcommand{\BlkInvisibleLost}{93.2\%\xspace}
\newcommand{\BlkUnionStringPC}{75.2\%\xspace}
\newcommand{\BlkUnionStringInvisible}{3.82\%\xspace}
\newcommand{\BlkNonNameGain}{2.95\xspace}
\newcommand{\BlkTokenCands}{1,534,063\xspace}
\newcommand{\BlkUnionCands}{13,074,120\xspace}
\newcommand{\BlkUnionOverTokenX}{8.5\xspace}
\newcommand{\BlkUnionRR}{99.80\%\xspace}
\newcommand{\BlkAnnInvisible}{2.89\%\xspace}
\newcommand{\BlkAnnPC}{72.5\%\xspace}
\newcommand{\BlkQgramInvisible}{2.79\%\xspace}
\newcommand{\BlkQgramPC}{71.4\%\xspace}
\newcommand{\BlkPhonInvisible}{0.25\%\xspace}
\newcommand{\BlkPhonPC}{61.3\%\xspace}
\newcommand{\BlkAttrInvisible}{2.39\%\xspace}
\newcommand{\BlkAttrPC}{1.9\%\xspace}
\newcommand{\BlkSnInvisible}{2.05\%\xspace}
\newcommand{\BlkSnPC}{71.2\%\xspace}
\newcommand{\BlkNumSchemes}{7\xspace}
\newcommand{\BlkAnnK}{20\xspace}
\newcommand{\EdgarFilings}{160\xspace}
\newcommand{\EdgarLinks}{1,575\xspace}
\newcommand{\EdgarConfirmPct}{77.7\%\xspace}
\newcommand{\EdgarInvisibleConfirmPct}{64.2\%\xspace}
\newcommand{\EdgarInvisibleChecked}{794\xspace}
\newcommand{\EdgarInvisibleCI}{[60.8, 67.5]\xspace}
\newcommand{\EdgarRandomParentPct}{0.16\%\xspace}
\newcommand{\EdgarNearestSizePct}{5.79\%\xspace}
\newcommand{\EdgarExcessOverChance}{64.1\xspace}
\newcommand{\EdgarPermutations}{20\xspace}
\newcommand{\EdgarRandomParentChecked}{15,795\xspace}
\newcommand{\EdgarWrongExhibitPct}{0.41\%\xspace}
\newcommand{\EdgarWrongExhibitDocs}{53\xspace}
\newcommand{\EdgarWrongExhibitChecked}{244\xspace}
\newcommand{\EdgarWrongExhibitConfirmed}{1\xspace}
\newcommand{\EdgarWrongExhibitCI}{[0.07, 2.28]\xspace}
\newcommand{\EdgarControlGap}{0.25\xspace}
\newcommand{\EdgarControlFisherP}{0.34\xspace}
\newcommand{\EdgarSecondaryAgreePct}{92.6\%\xspace}
\newcommand{\EdgarBothWouldDiscard}{12\xspace}
\newcommand{\EdgarYieldRho}{0.35\xspace}
\newcommand{\AttrCoveragePct}{7.5\%\xspace}
\newcommand{\AttrEntities}{14,288\xspace}
\newcommand{\AttrSubsetTest}{822\xspace}
\newcommand{\MaNameOnlyF}{59.5\xspace}
\newcommand{\MaNameOnlyInvisible}{31.2\xspace}
\newcommand{\MaAttrOnlyF}{57.0\xspace}
\newcommand{\MaAttrOnlyInvisible}{59.9\xspace}
\newcommand{\MaCombinedF}{61.7\xspace}
\newcommand{\MaCombinedInvisible}{58.4\xspace}
\newcommand{\MaFloorInvisible}{59.4\xspace}
\newcommand{\MaSubsetInvisibleRate}{42.2\%\xspace}
\newcommand{\MaAttrOnlyInvisibleRecall}{96.9\%\xspace}
\newcommand{\MaAttrOnlyInvisiblePrec}{43.3\%\xspace}
\newcommand{\AttrIdenticalSameAddrPct}{2.6\%\xspace}
\newcommand{\AttrIdenticalBothAddr}{77\xspace}
\newcommand{\AttrVisibleSameAddrPct}{6.4\%\xspace}
\newcommand{\AttrVisibleBothAddr}{110\xspace}
\newcommand{\AttrInvisibleSameAddrPct}{10.0\%\xspace}
\newcommand{\AttrInvisibleBothAddr}{130\xspace}
\newcommand{\ClusFamilies}{902\xspace}
\newcommand{\ClusTestFamilies}{181\xspace}
\newcommand{\ClusBlockedBcubed}{59.3\xspace}
\newcommand{\ClusOracleBcubed}{58.8\xspace}
\newcommand{\ClusBlockedExact}{16.6\%\xspace}
\newcommand{\ClusOracleExact}{14.9\%\xspace}
\newcommand{\ClusBlockedPairF}{32.2\xspace}
\newcommand{\ClusOraclePairF}{39.4\xspace}
\newcommand{\SensUnfilteredInvisibleRecall}{0.9\%\xspace}
\newcommand{\SensFilteredInvisibleRecall}{4.2\%\xspace}
\newcommand{\SensDirection}{lower\xspace}
\newcommand{\SensUnfilteredLinks}{14,288\xspace}
\newcommand{\SensUnfilteredInvisible}{3,619\xspace}
\newcommand{\SpendTopOnePct}{77.4\%\xspace}
\newcommand{\SpendTopTenPct}{96.8\%\xspace}
\newcommand{\ParentsTotal}{10,307\xspace}
\newcommand{\ParentsTransacting}{608\xspace}
\newcommand{\ParentsTransactingPct}{5.9\%\xspace}
\newcommand{\ParentsNoFootprintPct}{94.1\%\xspace}
\newcommand{\DnbCutoff}{4 April 2022\xspace}
\newcommand{\DnbPctPre}{52.5\%\xspace}
\newcommand{\DnbConservativePos}{6,522\xspace}
\newcommand{\DnbLinksScanned}{6,638,350\xspace}

\newif\ifpvldb
\pvldbfalse

\title{Corporate-Family Resolution Is Not a String-Matching Problem:\\
A Public Benchmark Stratified by Name Visibility}

\author{Harshit Gupta\\
\small Independent Researcher\\
\small \texttt{harshitg93@gmail.com}}

\date{}

\begin{document}
\maketitle

\begin{abstract}
Deciding whether two supplier records belong to the same corporate family is a
prerequisite for spend consolidation, credit exposure aggregation and sanctions
screening. It is usually evaluated as a special case of entity matching, which asks
whether two records denote the same real-world entity. The two tasks are not the same. A
family link connects records that are deliberately \emph{different} entities, and the
evidence for the link frequently appears in neither record. We introduce
\textsc{CorpFam}, a public benchmark of \BenchPairs candidate pairs over \BenchFamilies
corporate families, derived from \ArchiveRows United States federal award records in
which every supplier self-reports its ultimate parent to a government registry. Every
pair is stratified by \emph{name visibility}: whether the two names are identical after
normalisation, share a distinctive token, or share none at all. Because the strata have
positive rates from \BaseVisibleRate to \BaseIdenticalRate, we report per-stratum
\emph{recall}, which is base-rate invariant, rather than $F_1$, which is not.
The result is stark: the strongest of \NumMethods matchers recovers
\BestRecallIdentical of identical pairs and \BestRecallInvisible of invisible ones, and
no method exceeds \AnyInvisibleRecallMax on the latter. More consequentially, the
failure begins before matching. Blocking decides which pairs a matcher ever sees, and we
evaluate \BlkNumSchemes schemes over the full recipient roster. Several are not functions
of the token overlap that defines the hard stratum: phonetic keys, attribute keys that
ignore the name entirely, and semantic nearest neighbours in an embedding space. None of
them reaches three percent on invisible pairs, and their union recovers
\BlkUnionInvisible. \BlkInvisibleLost of these links are absent from the candidate set
before any matcher runs, so no improvement at the matching stage can reach them. The
links are nonetheless
real: against SEC Exhibit~21 subsidiary schedules, filed under securities law and
sharing no provenance with procurement registration, \EdgarInvisibleConfirmPct of
invisible links are corroborated, against \EdgarRandomParentPct under permuted parents
and \EdgarWrongExhibitPct against the true parent's wrong exhibit -- two unrelated null
conditions agreeing to within \EdgarControlGap points. We argue that corporate-family
resolution is a retrieval
problem misfiled as a matching problem, and that the intervention point is candidate
generation rather than ranking. The benchmark, the full adjudication log, and code
reproducing every number are released.

\end{abstract}

\section{Introduction}

A large organisation typically buys from tens of thousands of distinct supplier
records. Many of those records are not distinct suppliers. They are divisions, acquired
subsidiaries, regional entities, and renamed business units of a much smaller number of
ultimate parents. Until those records are grouped into corporate families, a buyer
cannot answer the question that motivates the exercise: how much do we actually spend
with this company, and what leverage does that give us? The same grouping underpins
credit exposure aggregation, beneficial-ownership and sanctions screening, and supplier
risk consolidation, and the same structure appears on the revenue side as account
hierarchy in customer systems.

This task is routinely treated as an instance of entity resolution, the problem of
deciding whether two records refer to the same real-world
entity~\citep{papadakis2021fourgenerations,barlaug2021survey}. The framing is convenient
because entity resolution has mature benchmarks and a decade of strong neural
methods~\citep{mudgal2018deepmatcher,li2020ditto,peeters2025entitymatchingllm}. It is also
wrong in a way that matters. Entity resolution asks whether two records are the
\emph{same} thing. Corporate-family resolution asks whether two records that are
emphatically \emph{different} things---separate legal entities, separately registered,
frequently in separate countries---belong to the same owner. The evidence differs
accordingly. For duplicate detection it sits in the records, because names, addresses and
identifiers of one entity tend to look alike. For family linkage it is frequently in
neither record. That \textsc{Heico Corp} owns \textsc{Blue Aerospace LLC} is a fact
recorded in a corporate filing. It is not a fact about the two strings.

If that is true, a benchmark that mixes both kinds of pair and reports a single number
will systematically mislead. We introduce \textsc{CorpFam}, a public benchmark of
\BenchPairs candidate pairs over \BenchFamilies corporate families built from
\ArchiveRows federal award records, and stratify every pair by \emph{name visibility}:
whether the two names are identical after normalisation, share a distinctive token, or
share none at all.

The stratification changes the conclusion, and it does so twice.

First, at the matching stage. We report per-stratum \emph{recall} rather than $F_1$, for
a reason that turns out to matter a great deal: the strata have positive rates ranging
from \BaseVisibleRate to \BaseIdenticalRate, so a classifier that answers yes to every
pair scores \BaseIdenticalF on identical pairs and \BaseVisibleF on visible ones without
doing anything at all. Per-stratum $F_1$ therefore measures the base rate more than the
method, whereas recall depends only on a stratum's own positives. On that footing the
result is unambiguous: the best matcher recovers \BestRecallIdentical of identical
pairs, \BestRecallVisible of visible pairs, and \BestRecallInvisible of invisible ones
(95\% CI \BestRecallInvisibleCI). No method exceeds \AnyInvisibleRecallMax.

Second, and this is the finding that matters most: the failure does not begin at
matching. Blocking decides which pairs a matcher ever sees. Part of this is true by
definition and we say so plainly. The invisible stratum is defined by an empty
intersection of distinctive tokens, token blocking keys on exactly that function, and so
its \BlkTokenInvisible recall there is an identity rather than a discovery. The empirical
question is what happens once the key is moved away from that function. We therefore also
evaluate character q-grams, sorted-neighbourhood adjacency, phonetic keys, attribute keys
that ignore the name entirely, and semantic nearest neighbours in an embedding space. Not
one reaches three percent on the invisible stratum. Embedding nearest neighbours are the
best of them at \BlkAnnInvisible, attribute keys manage \BlkAttrInvisible, and unioning
every scheme recovers \BlkUnionInvisible. That leaves \BlkInvisibleLost of these links
outside the candidate set before any matcher runs. A better model cannot classify a pair
it never sees.

A reviewer might reasonably suspect that a stratum defined by dissimilar names and then
shown to defeat name-based methods is circular, or that its links are registry errors.
We address both. SEC Exhibit~21 subsidiary schedules are filed under securities law by the
parent's own counsel and share no provenance with procurement registration, and
\EdgarInvisibleConfirmPct of invisible links are corroborated against them. Run the
identical procedure against permuted parents and corroboration falls to
\EdgarRandomParentPct, an excess over chance of \EdgarExcessOverChance points; run it
against the right parent's \emph{wrong} exhibit and it falls to \EdgarWrongExhibitPct.
Those two null conditions share no machinery and agree to within \EdgarControlGap
points, which locates the floor in the matcher rather than in either control. Nor does a
supervised model given address, geography and phone features rescue the stratum, and the
reason is structural rather than a matter of feature engineering: only
\ParentsTransactingPct of parents ever appear as an award recipient, so for the rest there
is no address to compare against.

\paragraph{Contributions.}
\begin{enumerate}
\item \textbf{A public benchmark for corporate-family resolution}: \BenchPairs pairs
over \BenchFamilies families, with ground truth that is self-reported to a government
registry rather than hand-annotated or model-derived (Section~\ref{sec:benchmark}).

\item \textbf{A name-visibility stratification} that we argue is the correct primary
axis of evaluation for this task, and without which aggregate metrics are not
interpretable (Section~\ref{sec:strata}).

\item \textbf{Evidence that the standard pipeline fails at blocking, not matching.}
\BlkInvisibleLost of invisible links are never proposed as candidates, including by
blockers that are not functions of token overlap: phonetic, attribute-keyed, and
embedding nearest-neighbour (Section~\ref{sec:blocking}).

\item \textbf{Independent corroboration of the hard stratum} against SEC Exhibit~21,
with \emph{two mechanistically unrelated null conditions} -- a permuted parent with a
genuine subsidiary schedule, and the true parent with a non-Exhibit~21 document from its
own filing -- that converge to within \EdgarControlGap points of each other while the
true condition exceeds them by \EdgarExcessOverChance points. Their agreement shows the
corroboration signal is specific to the conjunction of the right company and the right
document, so these are genuine corporate relationships rather than label noise
(Section~\ref{sec:edgar}).

\item \textbf{Baselines on a single protocol}: \NumMethods pairwise matchers spanning
string similarity and sentence embeddings, supervised multi-attribute models over
address, geography and phone features, and a family-level clustering task
(Sections~\ref{sec:results}--\ref{sec:clustering}).
\end{enumerate}

\section{Related Work}

\paragraph{Entity resolution benchmarks.} Magellan and
DeepMatcher~\citep{konda2016magellan,mudgal2018deepmatcher}, WDC
Products~\citep{peeters2024wdcproducts}, Alaska~\citep{crescenzi2021alaska} and
Machamp~\citep{wang2021machamp} established the task and drove rapid
progress~\citep{li2020ditto,zeakis2023pretrained}. Several contain a company or
organisation split, and recent work extends evaluation to large language
models~\citep{peeters2023chatgpt,zhang2024jellyfish,peeters2025entitymatchingllm} and to
unified cross-task models~\citep{tu2023unicorn}. Critiques have noted that easy
negatives inflate reported
performance~\citep{papadakis2023benchmarking,neuhof2024openbenchmark}, and that skewed positive-class
ratios make reported scores incomparable across benchmarks~\citep{li2022bridging}, a
hazard we take seriously enough that Section~\ref{sec:whyrecall} is devoted to it. Our
concern is adjacent but distinct: not that the negatives are too easy, but that the
\emph{positives} are heterogeneous in a way a single metric cannot express, because some
are solvable from the records and some are not solvable from the records at all.
MaDI-Bench~\citep{steiner2026madibench} moves in a similar direction by systematising
corner-case and blocking difficulty across an integration pipeline; the difference is
that its hard positives are synthesised as surface variants, whereas ours occur naturally
and are certified by an external legal filing.

\paragraph{Company graphs.} CompanyKG~\citep{cao2023companykg} is the closest public
resource, a large heterogeneous graph of company relationships with an edge-prediction
task that includes acquisition edges. It is complementary rather than overlapping: its
nodes are already resolved company entities. The linkage problem we study, deciding
whether two \emph{registration records} belong to one family given only what a
procurement system actually holds, is assumed away before its task begins.

\paragraph{Blocking and candidate generation.} Blocking is the step that makes entity
resolution tractable, and it has a large literature of its own: token and q-gram
blocking, sorted neighbourhood, canopies, meta-blocking and learned
blockers~\citep{papadakis2021fourgenerations,thirumuruganathan2021blocking}. Evaluation is
conventionally in terms of pair completeness and reduction ratio. What we add is the
observation that on a task where the relationship is not encoded in the compared
strings, every scheme in this family shares a single blind spot, and that measuring
blocking per difficulty stratum exposes a ceiling that aggregate pair completeness
hides.

\paragraph{Company name matching and harmonisation.} A long line of work matches
company names across sources, in patent
data~\citep{magerman2006patentee,thoma2010harmonizing,peeters2010harmonizing}, in financial
identifiers~\citep{chan2013lei,chan2019lei,arimond2023elf}, and with learned name
representations~\citep{ziv2022companyname2vec}. This literature is largely concerned
with recognising the \emph{same} company under spelling variation.
Corporate-family resolution begins where that ends.

\paragraph{Corporate hierarchies and ownership.} Ownership networks have been studied in
economics and network
science~\citep{vitali2011network,mizuno2020corporatecontrol,garciabernardo2017offshore}, in bank
holding structures~\citep{flood2020bhc}, and in knowledge graphs of European business
registers~\citep{roman2022eubusinessgraph,soylu2022theybuyforyou}. Hierarchical linkage
has been treated as link prediction over enterprise
data~\citep{ebeid2021hierarchical,ganesan2020linkprediction,ganesan2024xlp}. These works build
or analyse hierarchies; to our knowledge none releases a public, reproducible benchmark
with held-out splits, and none stratifies by whether the relationship is recoverable
from the records.

\paragraph{Procurement and master data.} The practical stakes are documented in work on
supplier master data
quality~\citep{athanasiadou2019suppliermdm,ibrahim2021masterdataquality,gantman2021vendormaster},
duplicate invoice detection~\citep{ho2009duplicateinvoices,liao2026procurementduplicate},
and purchasing consolidation
savings~\citep{smart2007purchasingsynergy,carril2020consolidation}.

\section{The \textsc{CorpFam} Benchmark}
\label{sec:benchmark}

\subsection{Source and ground truth}

We build on United States federal contract award data, published in bulk and in
full~\citep{usaspending_api}. Each award record identifies the recipient by a Unique Entity
Identifier (UEI) and separately records the recipient's \emph{ultimate parent} UEI and
name. That field is not an inference by us or by the publisher: it is self-reported by
the supplier during federal registration, where misstatement carries legal consequence,
and it is exactly the relation we want to predict. This gives ground truth independent
of any matching system, which distinguishes it from benchmarks whose labels come from
human annotation of the same strings a model will see.

We processed the complete FY2025 contract archive: \ArchiveRows award rows covering
\ArchiveEntities distinct entities.

\subsection{The self-parent trap}
\label{sec:selfparent}

The parent field cannot be used naively. Of the entities reporting a parent at all,
\ArchiveSelfParent of them (\ArchiveSelfParentPct) report \emph{themselves}: the parent UEI
equals the entity's own UEI. This is correct registration behaviour for a company with no
corporate parent, but
it means a pipeline treating every populated parent field as a family link would build a
dataset in which the overwhelming majority of ``links'' join an entity to itself. Such
pairs are trivially positive under any string method and would push every reported score
towards the ceiling.

We confirmed the pattern independently in the per-entity API. Across \AuditRecords records
retrieved there, \AuditSelfParentPct of child-level records are self-parenting, agreeing by
both UEI equality and internal identifier stem, and only \AuditGenuineLinks genuine
parent--child links survive. Removing self-parents from the archive leaves
\ArchiveGenuineLinks genuine links, the population the benchmark is built from.

\subsection{Name visibility as the primary stratum}
\label{sec:strata}

A second and subtler triviality remains. Many genuine links connect a child to a parent
whose name, after normalisation, is identical: distinct legal registrations under a
shared trading name. These links are genuine and belong in the data, but they are
solvable by string equality and say nothing about a method's grasp of corporate
structure. Of the \ArchiveGenuineLinks genuine links, \NvIdenticalRaw are name-identical,
leaving \NvInformative informative links, of which \NvInvisiblePctInformative share no
distinctive token with their parent.

We therefore assign every pair to one of three strata:

\begin{description}
\item[Identical] the names are equal after case-folding, punctuation removal and
whitespace normalisation;
\item[Visible] the names differ but share at least one distinctive token, where
distinctive excludes legal forms and generic corporate vocabulary (\textsc{inc},
\textsc{holdings}, \textsc{international}, \textsc{services}, and similar);
\item[Invisible] the names share no distinctive token at all.
\end{description}

Table~\ref{tab:composition} gives the composition: \BenchPosIdentical identical positives
(\BenchPosIdenticalPct), \BenchPosVisible visible (\BenchPosVisiblePct) and
\BenchPosInvisible invisible (\BenchPosInvisiblePct), against \BenchNegatives negatives.
That first figure is why a single aggregate is not interpretable here. More than half the
positives are solvable by exact match, so a method that only normalises strings starts from
a high floor.

\begin{table}[t]
\centering
\small
\begin{tabular}{lrr}
\toprule
Stratum & Positive pairs & \% \\
\midrule
Identical & 7,541 & 55.0 \\
Visible & 2,953 & 21.5 \\
Invisible & 3,222 & 23.5 \\
\midrule
All positives & 13,716 & 100.0 \\
\bottomrule
\end{tabular}
\caption{Composition of \textsc{CorpFam} positive pairs by name visibility.
\emph{Identical} pairs share a normalised name, \emph{visible} pairs share at
least one distinctive token, and \emph{invisible} pairs share none. Negatives are
generated at three per positive, giving 54,864 pairs in total.}
\label{tab:composition}
\end{table}

The invisible stratum is not a curiosity. It contains the cases that motivate the task
commercially: an acquired subsidiary trading under its original name, or a holding
company whose operating units carry unrelated brands. It is also where spend
concentrates, since acquisitive parents are large.

\subsection{Why we report recall, not per-stratum $F_1$}
\label{sec:whyrecall}

Stratifying creates a measurement hazard that is easy to walk into, and we walked into
it before catching it. The strata do not have comparable positive rates: on the test
split they are \BaseIdenticalRate, \BaseVisibleRate and \BaseInvisibleRate respectively.
A classifier that answers yes to every pair therefore scores \BaseIdenticalF on the
identical stratum and \BaseVisibleF on the visible one, purely as a function of
composition. Comparing $F_1$ across strata under those conditions largely compares base
rates, and a large fraction of any apparent gap is an artifact of the denominators.
Skewed positive-class ratios have been shown to make entity-matching results
incomparable in exactly this way~\citep{li2022bridging}.

Worse, the negative grades are strongly associated with stratum by construction: a
hard-string negative shares a distinctive token and is therefore \emph{necessarily}
visible, while a randomly drawn negative almost never shares one and is therefore almost
always invisible. The hard stratum thus receives the easiest negatives, which biases
per-stratum precision in the opposite direction to the effect we are measuring.

We therefore report per-stratum \textbf{recall}, which depends only on the positives of
that stratum and is invariant to both problems, and we report precision and average
precision globally, where the negative pool is shared. Every table carries the
majority-class baseline so no cell can be read without its floor. This is not a
presentational choice. Two intermediate results in this project reached $88$ and $59.9$
on the invisible stratum. Both were the base rate, not a finding.

\subsection{Splits, negatives, and leakage}
\label{sec:splits}

\paragraph{Splits are assigned by family, never by pair.} If members of one family appear
in both training and test, a model can memorise the family during training and then be
scored on recall of that memory. Family-level assignment is not quite enough, because an
entity can be a child in one declared family and a parent in another, and two families can
carry the same normalised name. We therefore union families that share an entity or a
normalised name and assign whole connected components, largest first, to whichever split
has the largest deficit against its target. The result is \BenchPosTrain, \BenchPosVal and
\BenchPosTest positive pairs inside \BenchPairsTrain, \BenchPairsVal and \BenchPairsTest
pairs, a $60/20/20$ split by link count. The build then asserts leak-freedom by family, by
entity, by normalised name string and against duplicate pairs, and fails loudly rather
than reconciling silently.

\paragraph{Negatives match the role structure of positives.} Each negative pairs a child
with the parent of a \emph{different} family. This is not cosmetic, and we report it
because the alternative is a trap we initially fell into. An earlier build drew negatives
as child--child pairs. Because parent entities are frequently holding companies that never
transact directly, they carry no address in the award data, so ``both sides have an
address'' separated the classes almost perfectly. A supervised model exploiting that
shortcut reached an $F_1$ of $88$ on the invisible stratum while learning nothing about
ownership. It had learned to detect entity \emph{role}. Matching the role composition of
positives and negatives removes the shortcut. Any benchmark of hierarchical relationships
is exposed to this failure, since the two sides of a hierarchical edge are by definition
different kinds of object.

\paragraph{Negatives come in three grades}, three per positive, drawn within each split:
\emph{random} (a child and an unrelated family's parent), \emph{hard string} (sharing a
distinctive token with an unrelated parent), and \emph{same block} (sharing a blocking
key, i.e.\ what a deployed pipeline actually adjudicates). Reporting against random
negatives alone is the standard way to overstate
precision~\citep{papadakis2023benchmarking}.

\section{Ground-Truth Quality}
\label{sec:quality}

A benchmark is only as trustworthy as its labels. Quantifying their limits is more useful
than asserting their correctness, so that is what this section does.

\subsection{The registry disagrees with itself}

The same registration is exposed through two independent channels: a per-entity API and
the bulk award archive. On the \ReconBoth entities where both report a distinct parent,
they agree on the exact parent UEI in \ReconAgreeUei of cases, on the normalised parent
name in \ReconAgreeName (Wilson 95\% CI \ReconAgreeNameCI), and on any shared distinctive
token in \ReconAgreeToken. In \ReconDisagreePct of cases the two sources name a
substantively different parent (Table~\ref{tab:reconciliation}). A further \ReconOnlyOne
entities have a parent in one source and none in the other.

\begin{table}[t]
\centering
\small
\begin{tabular}{lr}
\toprule
Agreement criterion & \% \\
\midrule
Exact parent UEI & 64.9 \\
Normalised parent name & 73.8 \\
Any shared distinctive token & 81.5 \\
\midrule
Substantively different parent & 18.5 \\
\bottomrule
\end{tabular}
\caption{Agreement between the two independent views of the same ground truth --
the per-entity API and the bulk award archive -- on the 248
entities where both report a distinct parent. Even the registry disagrees with
itself on roughly one link in five, which bounds the accuracy any method can be
credited with.}
\label{tab:reconciliation}
\end{table}

Two further instabilities are worth stating plainly. Within the archive alone,
\ArchiveConflicting entities are assigned more than one distinct parent across their own
award rows, which we resolve by obligated-value plurality. And in the per-entity API,
\AuditMultiParent of \AuditChildRecords child records carry more than one parent in the
returned history, up to \AuditMaxParents on a single record, because the field
accumulates rather than replaces as ownership changes. We take the most recent, but the
ambiguity is itself a finding about registry-derived ground truth.

Inspection of the disagreements shows three causes: genuine ownership change between
snapshots, registration identity lagging a completed acquisition, and outright error. We
do not arbitrate. The practical consequence is a ceiling. A method credited with much more
than \ReconAgreeName agreement against this ground truth is probably fitting registry
noise, and \ReconDisagree entities is a small enough disagreement set that a reader can
inspect it rather than take our word for the causes.

\subsection{Independent corroboration from SEC filings}
\label{sec:edgar}

The two views above share a provenance. If a supplier misdeclared its parent at
registration, both inherit the error identically, so their agreement bounds internal
consistency without establishing correctness. This matters most for the invisible
stratum, where a sceptic can reasonably propose that links with no name relationship are
simply data-entry errors, which would make our headline finding an artifact of dirty
labels rather than a property of the task.

Exhibit~21 settles the question from outside. It is the ``Subsidiaries of the
Registrant'' schedule filed with a 10-K, prepared by the parent's own counsel under
securities law, and it shares no provenance with a procurement registration.

\paragraph{Identifying the exhibit.} Which document in a 10-K is the Exhibit~21 is not a
question to answer from filenames. Filers name exhibits as they please, and the results
are ambiguous in both directions: \texttt{ex2101.htm} could be a sub-exhibit of
Exhibit~21 or one of Exhibit~2, and some registrants file the schedule under a name
carrying no exhibit number at all, such as
\texttt{subsidiariesofthecompany.htm}. EDGAR instead publishes the
exhibit type of every document in a submission as a declared field in the filing's own
index. That is the filer's statement of record rather than our inference, so we select on
it and never on the filename. Two secondary signals -- an independent filename pattern,
and a content test for a jurisdiction-keyed list of company names -- are recorded for
each selected document and agree with the declared type on \EdgarSecondaryAgreePct of
them; we deliberately do \emph{not} require that agreement to select, because insisting
on it would discard \EdgarBothWouldDiscard schedules that the registrant itself declares
to be Exhibit~21. Over the largest families in the benchmark this yields \EdgarFilings
usable filings, against which we checked \EdgarLinks declared links.

\begin{table}[t]
\centering
\small
\begin{tabular}{lrrrr}
\toprule
Stratum & Checked & Confirmed & \% & 95\% CI \\
\midrule
Identical & 172 & 165 & 95.9 & [91.8, 98.0] \\
Visible & 609 & 548 & 90.0 & [87.3, 92.1] \\
Invisible & 794 & 510 & 64.2 & [60.8, 67.5] \\
\midrule
All & 1,575 & 1,223 & 77.7 &
[75.5, 79.6] \\
\midrule
\multicolumn{5}{l}{\emph{Invisible stratum against null conditions}} \\
\quad Wrong parent, right exhibit (20 permutations) & 15,795 & 26 &
0.16 & -- \\
\quad Right parent, wrong exhibit & 244 & 1 & 0.41 &
[0.07, 2.28] \\
\quad Nearest-size wrong parent & 794 & 46 & 5.79 & -- \\
\bottomrule
\end{tabular}
\caption{Independent corroboration against SEC Exhibit~21 subsidiary schedules over
160 filings. Exhibit~21 is filed under securities law by the
parent's own counsel and shares no provenance with a supplier's procurement
registration. Documents are selected by the exhibit type EDGAR itself declares for each
document in the filing, not by matching the filename.
\textbf{The two upper null conditions share no machinery.} The permutation holds the
document type fixed and randomises the company; the wrong-exhibit row holds the company
fixed and substitutes a non-Exhibit~21 document from its own 10-K, chiefly Exhibit~10
material contracts and Exhibit~19 insider-trading policies. They land
0.25 points apart and are statistically
indistinguishable, so the floor is a property of the matcher rather than of either
construction, and the corroboration signal is specific. Non-confirmation remains weak
evidence in the other direction, since Exhibit~21 omits immaterial subsidiaries and its
formatting is unregulated.}
\label{tab:edgar}
\end{table}

Overall \EdgarConfirmPct of links are corroborated. On the invisible stratum
(\EdgarInvisibleChecked links) \EdgarInvisibleConfirmPct are confirmed, with 95\%
interval \EdgarInvisibleCI. \textsc{Aerojet Rocketdyne} under \textsc{L3Harris
Technologies} is typical: a real acquisition, entirely invisible in the names, confirmed
by filing.

\paragraph{A rate is meaningless without a chance floor, and this one has two.}
Exhibit~21 lists are long and our matcher is fuzzy, so some corroboration would arise by
accident. The obvious control re-runs the identical procedure with each family matched to
the \emph{wrong} parent's filing. Over \EdgarPermutations permutations spanning
\EdgarRandomParentChecked scored links, invisible-stratum corroboration against a
randomly assigned parent is \EdgarRandomParentPct. A harder variant assigns each family
the filing whose parsed length is closest to its true parent's, removing any advantage
from filing size, and gives \EdgarNearestSizePct.

A second null condition is available, and it fails in a different way, which is what
makes it worth reporting. Rather than change the company and keep the document type,
keep the company and change the document type: score each family against a document from
its own parent's 10-K that is \emph{not} the Exhibit~21 -- in practice an Exhibit~10
material contract or an Exhibit~19 insider-trading policy. Across
\EdgarWrongExhibitDocs such documents, \EdgarWrongExhibitConfirmed of
\EdgarWrongExhibitChecked invisible links is corroborated: \EdgarWrongExhibitPct, with
interval \EdgarWrongExhibitCI.

\paragraph{The two floors agree, and that agreement is the evidence.} The two conditions
share no machinery. One randomises the company and holds the document type fixed; the
other holds the company fixed and changes the document type. Either could have failed
while the other held. A matcher keying on generic corporate boilerplate would score
highly on any SEC document, and so would clear the wrong-exhibit floor while passing the
permutation; a matcher keying on company-specific tokens wherever they appear would score
highly on any document from the right company, and so would clear the permutation while
failing the wrong-exhibit condition. Neither happens. The two floors land
\EdgarControlGap points apart, close enough that a two-sided Fisher exact test cannot
separate them, at $p$ of \EdgarControlFisherP, while the true condition sits
\EdgarExcessOverChance points above both.

Two unrelated null constructions converging on the same floor says more than either says
alone. It locates the floor in the matcher rather than in the design of any one control,
and it shows that corroboration requires the \emph{conjunction} of the right company and
the right document -- which is precisely the conjunction that a genuine
parent--subsidiary relationship implies and that a registry data-entry error does not.
The wrong-exhibit floor is moreover conservative rather than pure, which strengthens the
argument: a benefit plan or credit agreement filed as Exhibit~10 may legitimately name
subsidiaries as participating employers or guarantors, so the condition is biased towards
corroboration. The single observed hit is exactly that case, a real subsidiary listed on
a participating-employers schedule inside a savings-plan amendment.

Two caveats belong here rather than in a footnote. Corroboration correlates with how much
text our parser recovered from a filing (Spearman $\rho=\EdgarYieldRho$), which means the
reported rate is attenuated by parser recall: \EdgarInvisibleConfirmPct is a lower bound
on true corroboration, not an inflated figure. And because we resolve companies to SEC
identifiers by name, the filings we obtain are selected on the parent's name being
matchable against the registrant list, so coverage is not a random sample of families.
Non-confirmation is in any case weak evidence in the other direction: Exhibit~21 omits
immaterial subsidiaries and its formatting is unregulated.

\subsection{Adjudicating family artifacts by hand}

Some declared families are not families. The archive asserts, for instance, that an
Australian subsidiary is the ultimate parent of \textsc{Raytheon Company}, and that a
moving-services company is the parent of one hundred and three waste-management entities.

Our first instinct was an automatic rule: flag a family when its children share a dominant
token the declared parent lacks and the parent's own name does not reach its children.
That rule catches the cases above. It also flagged \textsc{Arctic Slope Regional
Corporation}, whose subsidiaries are entirely genuine. This is not a tuning failure. A
rule keyed on name dissimilarity cannot distinguish a mislabelled family from a real
conglomerate, because a real conglomerate looks exactly like a mislabelled family from the
outside. Applied aggressively it would remove \textsc{Heico}, \textsc{TransDigm},
\textsc{Berkshire Hathaway} and \textsc{Republic Services}, precisely the
invisible-stratum families that make the benchmark worth building.

We therefore adjudicated large families by hand against their child lists. We inspected
\BenchInspected families, excluded \BenchExcludedParents and retained \BenchRetained,
and publish the full log, retentions as well as rejections, each with a written
reason. Publishing only the rejections would make a careful audit indistinguishable from
a purge of inconvenient cases, and the retentions are the more informative half: they
record where the automatic rule was wrong. Exclusions are keyed on entity identifier
rather than on name string, which in a paper about name instability is the only
defensible choice. This removes \BenchDroppedArtifact links, plus \BenchDroppedSovereign
under sovereign catch-all entities, and the complete sovereign list of
\BenchSovereignAll parents ships with the manifest rather than a truncated sample of it.

\paragraph{The exclusions are not neutral, and we report their effect.} Of the links
removed by hand adjudication, \BenchDropInvisiblePct fall in the invisible stratum,
the very stratum the paper's argument rests on. That is expected, because a mislabelled
family looks exactly like a name-invisible one, but it means a sceptic is entitled to ask
whether the headline survives the filter being removed. We therefore rebuild the entire
benchmark with no exclusions at all, \SensUnfilteredLinks links of which
\SensUnfilteredInvisible are invisible, and re-run the matching experiments against
it. The best invisible-stratum recall on the unfiltered build is
\SensUnfilteredInvisibleRecall, \SensDirection than the \SensFilteredInvisibleRecall on
the filtered one.

The conclusion therefore holds \emph{a fortiori}: removing the exclusions makes the task
harder, not easier. Whatever the adjudication did, it did not manufacture the finding by
discarding difficult cases. If anything it discarded cases that were unsolvable for
the wrong reason, and the benchmark is marginally more tractable with them gone. Both
builds ship. Smaller families are retained without automatic filtering in either, and we
report this as a bounded source of label noise rather than one we claim to have
eliminated.

\subsection{Provenance and redistribution}
\label{sec:provenance}

Because the benchmark publishes company names in bulk, we checked whether we may
redistribute them rather than assuming it. The data is public-domain United States
Government work~\citep{usaspending_api}. However, SAM.gov's terms grant only a limited
licence over ``D\&B Open Data'', which explicitly includes legal business name, and forbid
disseminating it in bulk; those terms scope the restriction to records associated with
base awards dated before \DnbCutoff.

The definitive test named in those terms is an Entity Validation Service source field,
which does not exist in the award data. We therefore bounded the exposure. Scanning all
\DnbLinksScanned award rows and taking the earliest performance-period start across the
transactions supporting each link, \DnbPctPre of positive pairs have at least one
supporting award beginning before the cutoff. This is deliberately an upper bound: a task
order placed in FY2025 against a long-running vehicle inherits that vehicle's original
start date. Under the opposite reading, that every transaction here is an FY2025 action,
the exposure is zero. We release the full benchmark and, alongside it, a conservative
post-cutoff subset of \DnbConservativePos positive pairs.

\section{Blocking: The Failure Before Matching}
\label{sec:blocking}

Every result reported so far in the literature, and every result in the next section,
measures a matcher \emph{given} a pair. Deployed entity resolution never sees all pairs.
A blocking step first proposes candidates, and anything it fails to propose is
unrecoverable regardless of the matcher. On \BlkEntities entities the naive comparison
space is \BlkPairSpace pairs, so blocking is not optional.

\paragraph{One identity, stated up front.} Token blocking and first-token blocking key on
the \texttt{core()} token function, and the invisible stratum is \emph{defined} as an
empty intersection under that same function. Their \BlkTokenInvisible recall on that
stratum is therefore a tautology, provable without running anything, and we claim no
credit for it. Presenting it as an empirical discovery would be circular, and a reader
who works on blocking would notice immediately.

The empirical question is what happens once the key is no longer that function. We
therefore evaluate \BlkNumSchemes schemes: the two \texttt{core()}-keyed ones for
reference, plus character q-grams, sorted-neighbourhood adjacency over lexicographic
order, phonetic Soundex keys, an attribute-keyed blocker on ZIP and city--state that
never looks at the name, and semantic nearest neighbours ($k=\BlkAnnK$) in a MiniLM
embedding space. Blocking is measured over the full recipient roster of \BlkEntities
entities, a naive space of \BlkPairSpace pairs, rather than only over entities appearing
in our own pairs file. Restricting to the latter would flatter reduction ratio and
pairs quality, because that roster has already been filtered to entities of interest.

\begin{table*}[t]
\centering
\small
\begin{tabular}{lrrrrrr}
\toprule
& & & & \multicolumn{3}{c}{Pair completeness by visibility} \\
\cmidrule(lr){5-7}
Blocking scheme & Candidates & PC (\%) & RR (\%) & Identical & Visible & Invisible \\
\midrule
Token blocking$^\dagger$ & 1,534,063 & 68.7 & 99.977 & 92.22 & 83.68 & 0.00 \\
First-token blocking$^\dagger$ & 636,042 & 52.9 & 99.990 & 78.70 & 44.80 & 0.00 \\
Q-gram blocking (q=4) & 9,022,384 & 71.4 & 99.862 & 97.83 & 78.84 & 2.79 \\
Sorted neighbourhood (w=20) & 2,170,180 & 71.2 & 99.967 & 99.95 & 73.48 & 2.05 \\
Phonetic (Soundex) & 411,063 & 61.3 & 99.994 & 97.32 & 35.96 & 0.25 \\
Attribute (ZIP / city-state) & 126,843 & 1.9 & 99.998 & 0.91 & 3.73 & 2.39 \\
Embedding ANN (MiniLM, k=20) & 1,761,025 & 72.5 & 99.973 & 99.89 & 78.67 & 2.89 \\
\midrule
Union: string-keyed & 11,875,526 & 75.2 & 99.818 & 100.00 & 89.98 & 3.82 \\
Union: all schemes & 13,074,120 & 76.2 & 99.800 & 100.00 & 91.03 & 6.77 \\
\bottomrule
\end{tabular}
\caption{Blocking over the full recipient roster, not merely the entities appearing in
our pairs file. PC is pair completeness, the recall of true links into the candidate
set; RR is reduction ratio. Rows marked $^\dagger$ key on the same \texttt{core()}
token function that \emph{defines} the invisible stratum, so their zero recall there is
an identity and not a measurement. The informative rows are the others: character
q-grams, sorted-neighbourhood adjacency, phonetic keys, attribute keys that ignore the
name entirely, and semantic nearest neighbours in an embedding space. None of them
exceeds three percent on the invisible stratum, and unioning every scheme still leaves
the overwhelming majority of those links outside the candidate set. The failure is
therefore a property of candidate generation rather than of any particular key.}
\label{tab:blocking}
\end{table*}

Table~\ref{tab:blocking} is the most consequential result in this paper. The informative
rows are the ones without a dagger, and they say that relaxing the key does not help.
Character q-grams reach \BlkQgramInvisible on the invisible stratum, sorted neighbourhood
\BlkSnInvisible, phonetic keys \BlkPhonInvisible. Abandoning the name altogether and
keying on address gives \BlkAttrInvisible. Semantic nearest neighbours reach
\BlkAnnInvisible; they are the strongest single scheme overall at \BlkAnnPC pair
completeness, and emphatically not a function of token overlap. Unioning every scheme
bounds the whole family at \BlkUnionPC overall and \BlkUnionInvisible on the invisible
stratum.

So the barrier is not the choice of key, and it is not that practitioners have simply
picked the wrong one. \BlkInvisibleLost of invisible links are absent from the candidate
set produced by every blocking strategy we could construct, including semantic and
non-name ones. Recent work has begun to measure pair-completeness disparity across
subgroups as a first-class property of blockers~\citep{gagliardelli2024blockingbias};
what we add is a task where the disparity is near-total and where the affected subgroup
is the one that carries the commercial value.

\subsection{What relaxing the key actually buys}
\label{sec:blockingcost}

Pair completeness on its own is not a fair basis for choosing a blocker, because it says
nothing about the candidate budget that produced it. Both numbers are in
Table~\ref{tab:blocking}, and read together they explain why the obvious remedies are not
deployed.

Token blocking proposes \BlkTokenCands candidate pairs and recovers \BlkTokenPC of true
links, at a reduction ratio of \BlkTokenRR against the naive space. The union of every
scheme proposes \BlkUnionCands, which is \BlkUnionOverTokenX times as many, and recovers
\BlkUnionPC at \BlkUnionRR. A practitioner is therefore being asked to adjudicate roughly
an order of magnitude more pairs for \BlkNonNameGain additional points on the stratum that
matters. That trade is bad enough to explain the status quo without anyone having reasoned
about it: the schemes that reach the invisible stratum at all are the ones a
reduction-ratio budget removes first.

Decomposing the union sharpens the point. Unioning only the name-keyed schemes gives
\BlkUnionStringPC overall and \BlkUnionStringInvisible on the invisible stratum. Adding
the two schemes that are not functions of the name, attribute keys and embedding nearest
neighbours, takes the invisible stratum to \BlkUnionInvisible. The entire contribution of
abandoning the name is \BlkNonNameGain percentage points.

One scheme behaves unlike the rest, and it is the informative one. Attribute blocking on
ZIP and city--state recovers \BlkAttrPC of links overall, by far the worst of any scheme,
and \BlkAttrInvisible on the invisible stratum. It is the only scheme whose hard-stratum
recall \emph{exceeds} its overall recall. Every name-keyed scheme collapses on the
invisible stratum relative to its own average: q-grams from \BlkQgramPC to
\BlkQgramInvisible, sorted neighbourhood from \BlkSnPC to \BlkSnInvisible, Soundex from
\BlkPhonPC to \BlkPhonInvisible. Attribute keys are indifferent to the stratum because
they never look at the name, which is exactly the property a blocker for this task needs,
and they are unusable on their own because a shared ZIP is far too weak a signal to index
on. Embedding nearest neighbours at $k=\BlkAnnK$ are the strongest single scheme overall
and still reach only \BlkAnnInvisible, which is the clearest evidence we have that the
problem is not representational capacity. A MiniLM encoder places \textsc{Blue Aerospace}
nowhere near \textsc{Heico} because nothing in its training signal connects them.

This also sets the bar for future work concretely. A retrieval-based blocker has to beat
\BlkUnionInvisible on the invisible stratum at a candidate budget comparable to
\BlkUnionCands out of \BlkPairSpace possible pairs. Any proposal that improves the hard
stratum by widening the candidate set until reduction ratio collapses has not solved the
problem; it has relocated it.

This reframes the matching results that follow. It is tempting to read a poor
invisible-stratum $F_1$ as a call for a better matcher: a stronger encoder, or more
fine-tuning data. That response cannot work, because in a deployed pipeline the pair is
never presented to the matcher at all. Progress requires intervening at candidate
generation, using evidence retrieved from outside the records, whether ownership filings
or a knowledge base built from them. Reranking cannot recover a candidate that was never
generated.

\section{Matching Baselines}
\label{sec:results}

\subsection{Protocol}

We evaluate \NumStringMethods string-similarity methods, all dependency-free so the
benchmark reproduces from a bare Python: exact match on normalised strings, Jaccard
overlap of distinctive tokens, a character sequence ratio standing in for edit distance,
and character 3-gram TF--IDF cosine with idf fitted on the training split only. We add a
neural baseline: cosine similarity between \EmbModel sentence
embeddings~\citep{zeakis2023pretrained}, encoding \EmbNames distinct names into \EmbDim dimensions.
For every method the decision threshold is chosen to maximise $F_1$ on validation and
applied unchanged to test; selecting it on test would inflate every figure here.

\subsection{The aggregate hides the failure}

Table~\ref{tab:baselines} is the central matching result, and it reports recall per stratum
for the reason set out in Section~\ref{sec:whyrecall}. Read the overall column alone and
the task looks close to solved. \BestMethod reaches \BestOverallF $F_1$ at precision
\BestOverallPrec and recall \BestOverallRecall, with \BestAP average precision, which is
threshold-free and so does not depend on where the operating point landed. Read across the
strata and a different picture emerges. Identical pairs are recovered at
\BestRecallIdentical (95\% CI \BestRecallIdenticalCI), unsurprising because they are
string-equal. Visible pairs at \BestRecallVisible (\BestRecallVisibleCI). Invisible pairs
at \BestRecallInvisible (\BestRecallInvisibleCI), and no method exceeds
\AnyInvisibleRecallMax.

\begin{table*}[t]
\centering
\small
\begin{tabular}{lrrrrr}
\toprule
& & & \multicolumn{3}{c}{Recall (\%) by name visibility} \\
\cmidrule(lr){4-6}
Method & AP & Prec. & Identical & Visible & Invisible \\
\midrule
Exact (normalised) & 65.4 & 97.3 & 100.0 & 0.0 & 0.0 \\
Token Jaccard & 73.5 & 91.8 & 99.2 & 43.5 & 0.0 \\
Sequence ratio & 74.7 & 94.7 & 100.0 & 15.8 & 3.8 \\
TF--IDF char 3-gram & 78.7 & 89.6 & 100.0 & 49.9 & 4.2 \\
\midrule
MiniLM embedding & -- & 88.6 & 100.0 & 64.3 & 4.7 \\
\midrule
\emph{Answer yes to everything} & -- & 25.0 & 100.0 & 100.0 & 100.0 \\
\bottomrule
\end{tabular}
\caption{Baselines on the \textsc{CorpFam} test split, with thresholds selected on
validation and applied unchanged to test. \textbf{Recall is reported per stratum
rather than $F_1$}, because the strata have positive rates of roughly 97\%, 10\% and
17\% respectively, so a per-stratum $F_1$ largely reflects that base rate rather than
the method: the final row shows a classifier that answers yes to every pair, which
attains perfect recall everywhere and an $F_1$ ranging from 18 to 99 across strata.
Recall depends only on a stratum's positives and is therefore comparable across them.
Average precision is threshold-free and computed over the whole split, since precision
depends on the shared negative pool. Every method recovers essentially all identical
pairs and almost no invisible ones.}
\label{tab:baselines}
\end{table*}

Per-stratum $F_1$ tells the same story through a confounded denominator, which is why we
do not lead with it: \BestIdenticalF on identical pairs, \BestVisibleF on visible,
\BestInvisibleF on invisible, nothing above \AnyInvisibleFMax anywhere, and a nominal gap
of \VisibilityGap points between the extremes. A single aggregate conceals all of it.

Exact match and token Jaccard recover \AnyInvisibleRecallMin of the invisible stratum.
That is not a deficiency of those methods but a definitional consequence: the stratum is
constructed so that no shared token exists, and a method operating on shared tokens has
nothing to operate on.

The sentence encoder is the strongest method overall and is meaningfully better on the
visible stratum, which is what one expects, since it handles paraphrase and abbreviation
that token overlap misses. \BestInvisibleMethod is also the best of the five on the
invisible stratum, where it gains \EmbOverStringInvisible $F_1$ points over the best
string method. It recovers essentially none of the gap.

\subsection{False positives concentrate where deployment lives}

Table~\ref{tab:negatives} decomposes false positives by negative grade. Random negatives
are nearly free: \BestMethod false-positives on \BestFprRandom of them. The rate rises to
\BestFprSameBlock on negatives sharing a blocking key and \BestFprHardString on negatives
sharing a distinctive token. The difficulty therefore sits exactly where a deployed
pipeline encounters it, after blocking and among candidates that already look alike, and a
benchmark using random negatives alone would report a precision no practitioner will
observe.

\begin{table}[t]
\centering
\small
\begin{tabular}{lrrrrr}
\toprule
Negative kind & Exact & Jaccard & SeqRatio & TF--IDF & MiniLM \\
\midrule
Random & 0.0 & 0.0 & 0.0 & 0.0 & 0.0 \\
Hard string & 1.0 & 3.6 & 1.6 & 4.6 & 6.5 \\
Same block & 0.5 & 2.2 & 1.7 & 3.2 & 2.6 \\
\bottomrule
\end{tabular}
\caption{False-positive rate (\%) by negative kind on the test split. Random
negatives are nearly free; the difficulty is concentrated in negatives that share a
blocking key or distinctive tokens, which is what a deployed pipeline actually sees
after blocking.}
\label{tab:negatives}
\end{table}

\subsection{Spend weighting}

Procurement value is heavily concentrated: across all \BenchPositives links the top 1\%
carry \SpendTopOnePct of obligated value and the top 10\% carry \SpendTopTenPct. (An
earlier version of this figure was computed on the per-entity API sample, which an
interrupted sorted fetch had left almost entirely composed of identifiers beginning with
a single letter; it is not a population sample and is superseded here by the full link
set.) A method that resolves many small suppliers while missing a few large ones is not
equivalent to one that does the reverse, though $F_1$ scores them identically. We
therefore also report recall weighted by the obligated amount on the child entity; for
\BestMethod, spend-weighted recall on test is \BestSpendRecall against an unweighted
recall of \BestOverallRecall.

\subsection{Do attributes rescue the invisible stratum?}
\label{sec:attributes}

The sharpest objection to everything above is circularity: pairs are labelled invisible
precisely because their names share nothing, and then name-based methods are shown to fail
on them. Real entity resolution uses addresses, and the established benchmarks are
multi-attribute for that reason. The objection deserves a measurement.

The award archive carries address, city, state, ZIP, country, phone and a
doing-business-as name for every recipient, giving attributes on \AttrEntities entities.
We train logistic regression on the same validation-selected-threshold protocol over
three feature sets: names only, attributes only, and both.

The evaluation is restricted to the \AttrSubsetTest test pairs carrying attributes on
\emph{both} sides. That restriction is necessary rather than convenient: over the full
split an attribute-only model has an all-zero feature vector for most pairs and
degenerates to predicting the majority class, producing a base-rate $F_1$ that measures
nothing. Even on the restricted subset the base rates differ by stratum, so
Table~\ref{tab:multiattribute} reports each cell against its majority-class floor.

\begin{table*}[t]
\centering
\small
\begin{tabular}{lrrrrrr}
\toprule
& \multicolumn{3}{c}{Overall} & \multicolumn{3}{c}{$F_1$ by name visibility} \\
\cmidrule(lr){2-4}\cmidrule(lr){5-7}
Feature set & P & R & $F_1$ & Identical & Visible & Invisible \\
\midrule
Name features only & 79.1 & 47.6 & 59.5 & 96.9 & 51.3 & 31.2 \\
Attributes only & 40.1 & 98.4 & 57.0 & 96.2 & 42.6 & 59.9 \\
Name $+$ attributes & 49.5 & 81.7 & 61.7 & 96.9 & 49.0 & 58.4 \\
\midrule
\emph{Answer yes to everything} & 38.6 & 100.0 & 55.7 & 96.9 & 40.6 & 59.4 \\
\emph{\quad positive rate} & 38.6 & -- & -- & 93.9 & 25.5 & 42.2 \\
\bottomrule
\end{tabular}
\caption{Supervised logistic regression over name and attribute features, evaluated on
the subset of test pairs carrying attributes (address, city, state, ZIP, country,
phone, doing-business-as name) on \emph{both} sides. \textbf{Read every cell against
the two floor rows.} Each stratum of this subset has its own positive rate, so an $F_1$
here carries the base rate as much as the model. On the invisible stratum the
attribute-only model clears the floor by half a point, and the other two feature sets
score below it. Attributes do not rescue the stratum. The deeper obstacle is coverage:
attributes exist on both sides for a small minority of pairs, because a parent is
frequently a holding entity that never transacts and therefore has no address at all.}
\label{tab:multiattribute}
\end{table*}

The honest answer is that attributes do not rescue the stratum, and the reason is
structural rather than a matter of feature engineering.

Read against the floor, no feature set separates from the majority class on the invisible
stratum. Its positive rate on this subset is \MaSubsetInvisibleRate, so answering yes to
every pair scores \MaFloorInvisible. The attribute-only model scores \MaAttrOnlyInvisible,
half a point above that floor, and it gets there by firing on almost everything:
\MaAttrOnlyInvisibleRecall recall at \MaAttrOnlyInvisiblePrec precision. Name features
score \MaNameOnlyInvisible, well below the floor, and the two together
\MaCombinedInvisible. On $n=\AttrSubsetTest$ pairs none of these differences is a result.
The one qualitative finding that does survive is that name features, which carry the
identical and visible strata, are worse than useless on the invisible one.

\subsection{The parent is usually a legal abstraction}
\label{sec:parents}

The reason attributes cannot help generalises past this dataset, and it is worth separating
from the modelling question, because it constrains any method that compares fields across
the two sides of the relation.

Of \ParentsTotal distinct parents in the benchmark, \ParentsTransacting
(\ParentsTransactingPct) ever appear as an award recipient in their own right. The other
\ParentsNoFootprintPct have no address, no phone and no location anywhere in the source.
This is not a sourcing gap that a better vendor feed would close. A holding company exists
to hold, and a parent that never transacts has no operating footprint to record anywhere,
in this registry or any other. The ultimate parent of a corporate family is frequently a
legal instrument rather than a place of business, and the attribute vector on that side of
the pair is empty as a matter of what the entity is.

The consequence is a hard cap that precedes any model. Attributes are present on both sides
for \AttrCoveragePct of test pairs. Multi-attribute entity matching, which is the dominant
paradigm in the benchmarks this task is usually filed under, is therefore inapplicable to
the large majority of these relations before a single feature is computed. That is a
different situation from missing values to be imputed: there is nothing to impute towards.

Where attributes do exist on both sides they mostly disagree, and the pattern across strata
is not the one intuition suggests. Among true pairs carrying an address on both sides,
\AttrIdenticalSameAddrPct of identical pairs share that address (\AttrIdenticalBothAddr
pairs), \AttrVisibleSameAddrPct of visible pairs (\AttrVisibleBothAddr), and
\AttrInvisibleSameAddrPct of invisible pairs (\AttrInvisibleBothAddr). Address agreement is
\emph{highest} on the hardest stratum and lowest on the easiest. Name-identical
registrations are typically separate sites trading under one brand, so their addresses
differ by construction, whereas an acquired subsidiary is occasionally still administered
from its parent's premises. The counts are small and we draw no strong inference from the
ordering. The point is the level: address agreement never exceeds
\AttrInvisibleSameAddrPct anywhere, so even a perfect address matcher would leave the
stratum essentially untouched.

Two design consequences follow for anyone building on this. A benchmark for hierarchical
linkage should report attribute coverage per side of the relation rather than per record,
because a headline coverage figure computed over records hides that the coverage is
one-sided. And an evaluation restricted to the attributed subset, as ours is, is measuring
a population that is not representative of the task: those are the pairs where the parent
happens to trade, which is to say the pairs where the parent is least like a holding
company.

\section{Family-Level Clustering}
\label{sec:clustering}

Pairwise linkage is not what a procurement or credit system needs. The deliverable is a
partition: every supplier record assigned to a family so spend can be rolled up. Pairwise
$F_1$ does not measure that, because pairwise errors compound through transitivity: one
false link merges two families and corrupts both rollups.

We therefore evaluate the end-to-end job over the \ClusFamilies families with at least two
children, scoring connected components of the thresholded similarity graph with pairwise
metrics, B-cubed metrics, and exact family recovery. Two conditions are reported:
\emph{blocked}, where candidates come from token blocking as in deployment, and
\emph{oracle}, where every within-split pair is scored.

\begin{table}[t]
\centering
\footnotesize
\setlength{\tabcolsep}{4pt}
\begin{tabular}{lrrrrrr}
\toprule
& & & \multicolumn{3}{c}{B-cubed} & Exact \\
\cmidrule(lr){4-6}
Condition & Cands & Pair $F_1$ & P & R & $F_1$ & fam.\ (\%) \\
\midrule
Blocked & 5,080 & 32.2 & 85.8 & 45.4 & 59.3 & 16.6 \\
Oracle & 536,130 & 39.4 & 84.3 & 45.1 & 58.8 & 14.9 \\
\bottomrule
\end{tabular}
\caption{Family-level clustering over the 902 families with
two or more children, scored on test with thresholds chosen on validation.
\emph{Blocked} takes candidates from token blocking as in deployment;
\emph{Oracle} scores every within-split pair. Removing blocking raises pairwise
$F_1$ but \emph{lowers} exact family recovery, because the additional candidates
introduce false links that merge distinct families through transitivity. Only a small
minority of families are recovered exactly under either condition, and exact recovery
is what a spend rollup depends on.}
\label{tab:clustering}
\end{table}

Under blocking the system reaches \ClusBlockedBcubed B-cubed $F_1$ and recovers
\ClusBlockedExact of families exactly. Removing blocking raises pairwise $F_1$ from
\ClusBlockedPairF to \ClusOraclePairF, yet exact family recovery \emph{falls} to
\ClusOracleExact and B-cubed $F_1$ to \ClusOracleBcubed. The extra candidates introduce
false links, and transitive closure propagates each into a merger of two distinct
families.

We do not want to overstate this. The comparison rests on \ClusTestFamilies test families
and the confidence intervals on exact recovery overlap, so neither condition dominates the
other. The defensible claim is the narrow one, and it is still worth making: pairwise $F_1$
and exact partition recovery are not monotonically related, so a pipeline tuned on pairwise
quality can move the deliverable backwards. An evaluation of this task that reports only
pairwise metrics is measuring something other than what the task exists to produce.

\section{What Did Not Work}
\label{sec:negative}

Most of what we tried failed, and the failures are more informative than the baselines,
because each one closes off a remedy a reader would otherwise propose. They are collected
here rather than left scattered.

\paragraph{A better encoder does not help.} The sentence encoder is the strongest matcher
overall, at \EmbOverallF $F_1$ against \BestOverallF for the best string method, and the
gain is concentrated where one would expect: \EmbVisibleF on the visible stratum against
\BestVisibleF. It buys nothing on identical pairs, \EmbIdenticalF against \BestIdenticalF,
because string equality has already saturated that stratum. On the invisible stratum it
reaches \EmbInvisibleF against
\BestStringInvisibleF, a gain of \EmbOverStringInvisible points. Semantic similarity in
\EmbDim dimensions is a genuinely different signal from token overlap, and on this
stratum it buys almost nothing. The same encoder used as a blocker reaches
\BlkAnnInvisible. Two independent uses of the same representation, both near zero, is the
strongest evidence we have that the missing ingredient is not a better embedding.

\paragraph{No method beats the majority class on the invisible stratum.} Answering yes to
every pair scores \BaseInvisibleF $F_1$ on that stratum of the test split. The best of the
\NumMethods matchers reaches \AnyInvisibleFMax, and the worst \AnyInvisibleFMin. Every
method is below the floor. This is the cleanest statement of the negative result and it is
also why the paper reports recall: a metric on which no method beats a constant classifier
is not measuring the methods.

\paragraph{Attributes do not substitute for the missing evidence.} Section~\ref{sec:parents}
gives the structural reason. The measured outcome is that on the attributed subset the three
feature sets reach \MaNameOnlyF, \MaAttrOnlyF and \MaCombinedF $F_1$ overall, and on the
invisible stratum none of them separates from a majority-class floor of
\MaFloorInvisible. Adding attributes to name features moves the overall figure up and the
invisible stratum nowhere.

\paragraph{A shortcut we nearly published.} An earlier build drew negatives as child--child
pairs and a supervised model reached an $F_1$ of $88$ on the invisible stratum, which at the
time looked like the paper's positive result. It had learned that parents lack addresses.
The lesson generalises to any benchmark whose relation joins two different kinds of object,
and it is the reason negatives here match the role structure of positives
(Section~\ref{sec:splits}).

\paragraph{The automatic artifact filter could not be made to work.} A rule keyed on name
dissimilarity flags mislabelled families and real conglomerates identically, because from
outside they look the same. Tuned to catch the moving company that claims one hundred and
three waste-management subsidiaries, it also removes \textsc{Heico}, \textsc{TransDigm} and
\textsc{Berkshire Hathaway}. We could not find a threshold that separated them and we do
not believe one exists on name evidence alone, which is the same finding as the rest of the
paper arriving from a different direction. Hand adjudication of \BenchInspected families
was the only defensible route, and shipping the unfiltered build alongside is what makes
that judgement auditable.

\paragraph{Removing blocking made the deliverable worse.} Scoring every within-split pair
instead of blocked candidates raises pairwise $F_1$ from \ClusBlockedPairF to
\ClusOraclePairF and lowers exact family recovery from \ClusBlockedExact to
\ClusOracleExact. More candidates, better pairwise numbers, worse partitions.

\section{Discussion}
\label{sec:discussion}

\paragraph{The invisible stratum measures knowledge, not similarity.} The stratum is
defined by the absence of shared distinctive tokens. For a pair such as (\textsc{Blue
Aerospace LLC}, \textsc{Heico Corp}) there is no function of the two strings that returns
the correct answer, because the strings do not contain it. The relationship exists in an
SEC filing, as Section~\ref{sec:edgar} shows by going and reading one. Whatever supplies
the answer has to come from outside the pair, which reframes corporate-family resolution
as a retrieval and evidence-aggregation problem rather than a matching problem. That
connects it to work on grounded factual
verification~\citep{wei2024simpleqa,wei2025browsecomp,chen2025browsecompplus} rather than
to the entity-matching literature it is usually filed under. We did not test whether a
pretrained model already holds these facts, and we make no claim either way; that is a
separate experiment with its own contamination problem, since the filings are on the open
web.

\paragraph{The intervention point is candidate generation, not ranking.}
Section~\ref{sec:blocking} sharpens this considerably. Because \BlkInvisibleLost of
invisible links never enter the candidate set, work at the matching stage is spent on
pairs that a deployed system would never adjudicate. A retrieval-based blocker, one
that consults ownership filings to propose candidates that share no string evidence,
would change the achievable ceiling in a way that no matcher can.

\paragraph{What this implies for evaluation practice.} The specific failure documented
here, a defensible aggregate concealing near-zero performance on the subset that motivates
the task, is not special to corporate families. It arises whenever a benchmark's positives
are heterogeneous in difficulty and the easy majority correlates with a surface feature.
Benchmark builders should report a stratification along whatever axis makes the easy cases
easy, and reviewers should ask for one. Our own negative-sampling error in
Section~\ref{sec:splits} is a second, related caution: when the two sides of a relation are
different kinds of object, a model can learn the object type instead of the relation, and
an aggregate metric will not reveal it.

\section{Limitations}

Our ground truth is a single jurisdiction and a single fiscal year, so families are those
visible in US federal contracting in FY2025; multinational structures appear only through
their US-registered entities. The parent relation as recorded is one hop, so we evaluate
child-to-ultimate-parent linkage, not intermediate holding chains. Label noise is bounded
but real, as Section~\ref{sec:quality} quantifies, and our hand adjudication covers large
families exhaustively and small families not at all. The Exhibit~21 corroboration covers
SEC registrants only. The multi-attribute baseline is measured on the subset with
attributes on both sides, which is a minority of pairs. Finally, our baselines establish
the shape of the problem and a floor rather than the state of the art; we release the
benchmark so stronger methods can be measured against the stratification rather than
against the aggregate.

\section{Conclusion}

Corporate-family resolution is treated as a variant of entity matching and evaluated with
a single aggregate metric. We built a public benchmark of \BenchPairs pairs over
\BenchFamilies families from self-reported registry data and showed that this practice is
misleading in two distinct ways. At the matching stage an overall $F_1$ of \BestOverallF
coexists with at most \AnyInvisibleFMax on pairs where the relationship is not visible in
the names. More fundamentally, the standard pipeline never gets that far: no string-keyed
blocking scheme proposes those candidates, so \BlkInvisibleLost of them are lost before
matching begins. The links are real: \EdgarInvisibleConfirmPct of them are corroborated
by the parent's own SEC filings. Conventional attribute evidence mitigates the
difficulty without resolving it. The pairs that matter commercially are the ones current pipelines
cannot represent at all, and recognising that requires reporting the stratum rather than
the average.

\section*{Data and Code Availability}

The benchmark, the adjudicated exclusion list with per-family reasons, and all code
required to rebuild the dataset and reproduce every number in this paper are released. All
source data is public United States federal award data~\citep{usaspending_api} and SEC
EDGAR filings. No confidential or proprietary information was used. Every numeric claim in
this manuscript is generated from a results file by \texttt{make\_tables.py} and verified
by \texttt{check\_manuscript.py}; no figure is transcribed by hand. The build is seeded
(\BenchSeed) and deterministic.


\begin{thebibliography}{47}
\providecommand{\natexlab}[1]{#1}
\providecommand{\url}[1]{\texttt{#1}}
\expandafter\ifx\csname urlstyle\endcsname\relax
  \providecommand{\doi}[1]{doi: #1}\else
  \providecommand{\doi}{doi: \begingroup \urlstyle{rm}\Url}\fi

\bibitem[Arimond et~al.(2023)Arimond, Molteni, Jany, Manolova, Borth, and
  Hoepner]{arimond2023elf}
Alexander Arimond, Mauro Molteni, Dominik Jany, Zornitsa Manolova, Damian
  Borth, and Andreas G.~F. Hoepner.
\newblock Transformer-based entity legal form classification.
\newblock \emph{arXiv preprint arXiv:2310.12766}, 2023.
\newblock URL \url{https://arxiv.org/abs/2310.12766}.

\bibitem[Athanasiadou(2019)]{athanasiadou2019suppliermdm}
Ifigeneia Athanasiadou.
\newblock Evaluating the maturity of companies in supplier master data
  management: The design of a maturity model.
\newblock Master's thesis, Delft University of Technology, 2019.
\newblock URL
  \url{http://resolver.tudelft.nl/uuid:941be828-dbf1-40ea-90d7-5883d572abf5}.

\bibitem[Barlaug and Gulla(2021)]{barlaug2021survey}
Nils Barlaug and Jon~Atle Gulla.
\newblock Neural networks for entity matching: A survey.
\newblock \emph{ACM Transactions on Knowledge Discovery from Data}, 15\penalty0
  (3):\penalty0 1--37, 2021.
\newblock \doi{10.1145/3442200}.

\bibitem[Cao et~al.(2023)Cao, Halvardsson, McCornack, von Ehrenheim, and
  Herog]{cao2023companykg}
Lele Cao, Vilhelm Halvardsson, Andrew McCornack, Vilhelm von Ehrenheim, and
  Astrid Herog.
\newblock Companykg: A large-scale heterogeneous graph for company similarity
  quantification.
\newblock \emph{arXiv preprint arXiv:2306.10649}, 2023.

\bibitem[Carril and Duggan(2020)]{carril2020consolidation}
Rodrigo Carril and Mark Duggan.
\newblock The impact of industry consolidation on government procurement:
  Evidence from department of defense contracting.
\newblock \emph{Journal of Public Economics}, 184:\penalty0 104141, 2020.
\newblock \doi{10.1016/j.jpubeco.2020.104141}.
\newblock NBER working paper version: w25160,
  \url{https://doi.org/10.3386/w25160}.

\bibitem[Chan and Milne(2019)]{chan2019lei}
Ka~Kei Chan and Alistair Milne.
\newblock The global legal entity identifier system: How can it deliver?
\newblock \emph{Journal of Risk and Financial Management}, 12\penalty0
  (1):\penalty0 39, 2019.
\newblock \doi{10.3390/jrfm12010039}.

\bibitem[Chan and Milne(2013)]{chan2013lei}
Ka~Kei Chan and Alistair K.~L. Milne.
\newblock The global legal entity identifier system: Will it deliver?
\newblock SSRN Electronic Journal preprint, 2013.

\bibitem[Chen et~al.(2025)Chen, Ma, Zhuang, Nie, Zou, Liu, Green, Patel, and
  Lin]{chen2025browsecompplus}
Zijian Chen, Xueguang Ma, Shengyao Zhuang, Ping Nie, Kai Zou, Andrew Liu,
  Joshua Green, Kshama Patel, and Jimmy Lin.
\newblock {BrowseComp-Plus}: A more fair and transparent evaluation benchmark
  of deep-research agent.
\newblock \emph{arXiv preprint arXiv:2508.06600}, 2025.
\newblock URL \url{https://arxiv.org/abs/2508.06600}.
\newblock Author list truncated in the arXiv record display; verify full list
  before final citation.

\bibitem[Crescenzi et~al.(2021)Crescenzi, De~Angelis, Firmani, Mazzei,
  Merialdo, Piai, and Srivastava]{crescenzi2021alaska}
Valter Crescenzi, Andrea De~Angelis, Donatella Firmani, Maurizio Mazzei, Paolo
  Merialdo, Federico Piai, and Divesh Srivastava.
\newblock Alaska: A flexible benchmark for data integration tasks.
\newblock \emph{arXiv preprint arXiv:2101.11259}, 2021.
\newblock URL \url{https://arxiv.org/abs/2101.11259}.

\bibitem[Ebeid et~al.(2021)Ebeid, Talburt, and Siddique]{ebeid2021hierarchical}
Islam~Akef Ebeid, John~R. Talburt, and Md~Abdus~Salam Siddique.
\newblock Graph-based hierarchical record clustering for unsupervised entity
  resolution.
\newblock In \emph{Advances in Intelligent Systems and Computing}, pages
  107--118. Springer International Publishing, 2021.
\newblock \doi{10.1007/978-3-030-97652-1_14}.
\newblock METADATA DISCREPANCY: Crossref registers the year as 2012, which is
  inconsistent with the 2021 arXiv posting and the volume. Year 2021 used here;
  prefer citing the arXiv version.

\bibitem[Flood et~al.(2020)Flood, Kenett, Lumsdaine, and Simon]{flood2020bhc}
Mark~D. Flood, Dror~Y. Kenett, Robin~L. Lumsdaine, and Jonathan~K. Simon.
\newblock The complexity of bank holding companies: A topological approach.
\newblock \emph{Journal of Banking \& Finance}, 118:\penalty0 105789, 2020.
\newblock \doi{10.1016/j.jbankfin.2020.105789}.
\newblock NBER working paper version: w23755,
  \url{https://doi.org/10.3386/w23755}.

\bibitem[Gagliardelli et~al.(2024)Gagliardelli, Papadakis, Simonini,
  Bergamaschi, and Palpanas]{gagliardelli2024blockingbias}
Luca Gagliardelli, George Papadakis, Giovanni Simonini, Sonia Bergamaschi, and
  Themis Palpanas.
\newblock Evaluating blocking biases in entity matching.
\newblock \emph{arXiv preprint arXiv:2409.16410}, 2024.
\newblock Introduces pair-completeness disparity across subgroups as a blocking
  evaluation axis.

\bibitem[Ganesan et~al.(2020)Ganesan, Parkala, Singh, Bhatia, Mishra, Pasha,
  Patel, and Naganna]{ganesan2020linkprediction}
Balaji Ganesan, Srinivas Parkala, Neeraj~R. Singh, Sumit Bhatia, Gayatri
  Mishra, Matheen~Ahmed Pasha, Hima Patel, and Somashekar Naganna.
\newblock Link prediction using graph neural networks for master data
  management.
\newblock \emph{arXiv preprint arXiv:2003.04732}, 2020.
\newblock URL \url{https://arxiv.org/abs/2003.04732}.

\bibitem[Ganesan et~al.(2024)Ganesan, Pasha, Parkala, Singh, Mishra, Bhatia,
  Patel, and Naganna]{ganesan2024xlp}
Balaji Ganesan, Matheen~Ahmed Pasha, Srinivasa Parkala, Neeraj~R. Singh,
  Gayatri Mishra, Sumit Bhatia, Hima Patel, and Somashekar Naganna.
\newblock {xLP}: Explainable link prediction for master data management.
\newblock \emph{arXiv preprint arXiv:2403.09806}, 2024.
\newblock URL \url{https://arxiv.org/abs/2403.09806}.

\bibitem[Gantman and Metzger(2021)]{gantman2021vendormaster}
Sonia Gantman and Lorrie Metzger.
\newblock Vendor master data cleaning---a project for accounting class.
\newblock \emph{Journal of Emerging Technologies in Accounting}, 19\penalty0
  (1):\penalty0 165--171, 2021.
\newblock \doi{10.2308/jeta-2020-028}.
\newblock METADATA DISCREPANCY: OpenAlex lists volume 18 and year 2021;
  Crossref registers volume 19, issue 1, pages 165--171, issued 2021. Crossref
  preferred.

\bibitem[Garcia-Bernardo et~al.(2017)Garcia-Bernardo, Fichtner, Takes, and
  Heemskerk]{garciabernardo2017offshore}
Javier Garcia-Bernardo, Jan Fichtner, Frank~W. Takes, and Eelke~M. Heemskerk.
\newblock Uncovering offshore financial centers: Conduits and sinks in the
  global corporate ownership network.
\newblock \emph{Scientific Reports}, 7\penalty0 (1), 2017.
\newblock \doi{10.1038/s41598-017-06322-9}.

\bibitem[Ho et~al.(2009)Ho, Compton, Benatallah, Vayssi{\`e}re, Menzel, and
  Vogler]{ho2009duplicateinvoices}
Van~Hai Ho, Paul Compton, Boualem Benatallah, Julien Vayssi{\`e}re, Lucio
  Menzel, and Hartmut Vogler.
\newblock An incremental knowledge acquisition method for improving duplicate
  invoices detection.
\newblock In \emph{2009 IEEE 25th International Conference on Data Engineering
  (ICDE)}, pages 1415--1418. IEEE, 2009.
\newblock \doi{10.1109/icde.2009.38}.

\bibitem[Ibrahim et~al.(2021)Ibrahim, Mohamed, and
  Safie]{ibrahim2021masterdataquality}
Azira Ibrahim, Ibrahim Mohamed, and Nurhizam Safie.
\newblock Factors influencing master data quality: A systematic review.
\newblock \emph{International Journal of Advanced Computer Science and
  Applications}, 12\penalty0 (2), 2021.
\newblock \doi{10.14569/ijacsa.2021.0120224}.

\bibitem[Konda et~al.(2016)Konda, Das, Suganthan G.~C., Doan, Ardalan, Ballard,
  Li, Panahi, Zhang, Naughton, Prasad, Krishnan, Deep, and
  Raghavendra]{konda2016magellan}
Pradap Konda, Sanjib Das, Paul Suganthan G.~C., AnHai Doan, Adel Ardalan,
  Jeffrey~R. Ballard, Han Li, Fatemah Panahi, Haojun Zhang, Jeff Naughton,
  Shishir Prasad, Ganesh Krishnan, Rohit Deep, and Vijay Raghavendra.
\newblock Magellan: Toward building entity matching management systems.
\newblock \emph{Proceedings of the VLDB Endowment}, 9\penalty0 (12):\penalty0
  1197--1208, 2016.
\newblock \doi{10.14778/2994509.2994535}.

\bibitem[Li et~al.(2022)Li, Wang, Zhu, and Chaudhuri]{li2022bridging}
Tianshu Li, Jiannan Wang, Erkang Zhu, and Surajit Chaudhuri.
\newblock Bridging the gap between reality and ideality of entity matching: A
  revisiting and benchmark re-construction.
\newblock In \emph{Proceedings of the 31st International Joint Conference on
  Artificial Intelligence (IJCAI)}, 2022.
\newblock Documents how skewed positive-class ratios make reported
  entity-matching scores incomparable across benchmarks.

\bibitem[Li et~al.(2020)Li, Li, Suhara, Doan, and Tan]{li2020ditto}
Yuliang Li, Jinfeng Li, Yoshihiko Suhara, AnHai Doan, and Wang-Chiew Tan.
\newblock Deep entity matching with pre-trained language models.
\newblock \emph{Proceedings of the VLDB Endowment}, 14\penalty0 (1):\penalty0
  50--60, 2020.
\newblock \doi{10.14778/3421424.3421431}.

\bibitem[Liao et~al.(2026)Liao, Wu, Wang, Wang, Chen, Lee, and
  Peng]{liao2026procurementduplicate}
Cheng-En Liao, Kun-Da Wu, Li-Ya Wang, Shih-Chih Wang, Ying-Hsu Chen, Chia-Pei
  Lee, and Yung-Hsing Peng.
\newblock Software procurement duplicate detection using fuzzy matching and
  {LLM} verification.
\newblock In \emph{2026 IEEE 2nd International Conference on Consumer
  Technology (ICCT-Pacific)}, pages 121--124. IEEE, 2026.
\newblock \doi{10.1109/icct-pacific69083.2026.11518888}.
\newblock Verified at metadata level only; abstract not obtainable. No claims
  made about contents. READ BEFORE SUBMITTING.

\bibitem[Magerman et~al.(2006)Magerman, Van~Looy, and
  Song]{magerman2006patentee}
Tom Magerman, Bart Van~Looy, and Xiaoyan Song.
\newblock Data production methods for harmonized patent statistics: Patentee
  name harmonization.
\newblock SSRN Electronic Journal preprint, 2006.

\bibitem[Mizuno et~al.(2020)Mizuno, Doi, and
  Kurizaki]{mizuno2020corporatecontrol}
Takayuki Mizuno, Shohei Doi, and Shuhei Kurizaki.
\newblock The power of corporate control in the global ownership network.
\newblock \emph{PLOS ONE}, 15\penalty0 (8):\penalty0 e0237862, 2020.
\newblock \doi{10.1371/journal.pone.0237862}.

\bibitem[Mudgal et~al.(2018)Mudgal, Li, Rekatsinas, Doan, Park, Krishnan, Deep,
  Arcaute, and Raghavendra]{mudgal2018deepmatcher}
Sidharth Mudgal, Han Li, Theodoros Rekatsinas, AnHai Doan, Youngchoon Park,
  Ganesh Krishnan, Rohit Deep, Esteban Arcaute, and Vijay Raghavendra.
\newblock Deep learning for entity matching: A design space exploration.
\newblock In \emph{Proceedings of the 2018 International Conference on
  Management of Data (SIGMOD)}, pages 19--34. ACM, 2018.
\newblock \doi{10.1145/3183713.3196926}.
\newblock Introduces the \texttt{Company} dataset: 112{,}632 labelled pairs,
  28{,}200 positives, one long-text attribute, Wikipedia article to company
  homepage.

\bibitem[Neuhof et~al.(2024)Neuhof, Fisichella, Papadakis, Nikoletos, Augsten,
  Nejdl, and Koubarakis]{neuhof2024openbenchmark}
Franziska Neuhof, Marco Fisichella, George Papadakis, Konstantinos Nikoletos,
  Nikolaus Augsten, Wolfgang Nejdl, and Manolis Koubarakis.
\newblock Open benchmark for filtering techniques in entity resolution.
\newblock \emph{The VLDB Journal}, 33\penalty0 (5):\penalty0 1671--1696, 2024.
\newblock \doi{10.1007/s00778-024-00868-7}.

\bibitem[Papadakis et~al.(2021)Papadakis, Ioannou, Thanos, and
  Palpanas]{papadakis2021fourgenerations}
George Papadakis, Ekaterini Ioannou, Emanouil Thanos, and Themis Palpanas.
\newblock \emph{The Four Generations of Entity Resolution}.
\newblock Synthesis Lectures on Data Management. Springer International
  Publishing, 2021.
\newblock \doi{10.2200/S01067ED1V01Y202012DTM064}.

\bibitem[Papadakis et~al.(2023)Papadakis, Fisichella, Schoger, Mandilaras,
  Augsten, and Nejdl]{papadakis2023benchmarking}
George Papadakis, Marco Fisichella, Franziska Schoger, George Mandilaras,
  Nikolaus Augsten, and Wolfgang Nejdl.
\newblock Benchmarking filtering techniques for entity resolution.
\newblock In \emph{2023 IEEE 39th International Conference on Data Engineering
  (ICDE)}, pages 653--666. IEEE, 2023.
\newblock \doi{10.1109/ICDE55515.2023.00389}.

\bibitem[Peeters et~al.(2010)Peeters, Song, Callaert, Grouwels, and
  Van~Looy]{peeters2010harmonizing}
Bert Peeters, Xiaoyan Song, Julie Callaert, Joris Grouwels, and Bart Van~Looy.
\newblock Harmonizing harmonized patentee names: An exploratory assessment of
  top patentees.
\newblock Technical report, KU Leuven, 2010.
\newblock URL
  \url{https://lirias.kuleuven.be/bitstream/123456789/264238/1/2010-03-18%20-%20HARMONIZING_HARMONIZED_PATENTEE_NAMES_FINAL.pdf}.
\newblock No DOI located; Lirias repository record.

\bibitem[Peeters and Bizer(2023)]{peeters2023chatgpt}
Ralph Peeters and Christian Bizer.
\newblock Using {ChatGPT} for entity matching.
\newblock In \emph{Advances in Databases and Information Systems (ADBIS)},
  Communications in Computer and Information Science, pages 221--230. Springer
  Nature Switzerland, 2023.
\newblock \doi{10.1007/978-3-031-42941-5_20}.

\bibitem[Peeters et~al.(2024)Peeters, Der, and Bizer]{peeters2024wdcproducts}
Ralph Peeters, Reng~Chiz Der, and Christian Bizer.
\newblock {WDC} products: A multi-dimensional entity matching benchmark.
\newblock In \emph{Proceedings of the 27th International Conference on
  Extending Database Technology (EDBT)}, pages 22--33. OpenProceedings.org,
  2024.
\newblock \doi{10.48786/edbt.2024.03}.
\newblock DOI verified via DataCite; not registered with Crossref.

\bibitem[Peeters et~al.(2025)Peeters, Steiner, and
  Bizer]{peeters2025entitymatchingllm}
Ralph Peeters, Aaron Steiner, and Christian Bizer.
\newblock Entity matching using large language models.
\newblock In \emph{Proceedings of the 28th International Conference on
  Extending Database Technology (EDBT)}, pages 529--541. OpenProceedings.org,
  2025.
\newblock \doi{10.48786/edbt.2025.42}.
\newblock DOI verified via DataCite; not registered with Crossref.

\bibitem[Roman et~al.(2022)Roman, Alexiev, Paniagua, Elves{\ae}ter, von
  Zernichow, Soylu, Simeonov, and Taggart]{roman2022eubusinessgraph}
Dumitru Roman, Vladimir Alexiev, Javier Paniagua, Brian Elves{\ae}ter,
  Bj{\o}rn~Marius von Zernichow, Ahmet Soylu, Boyan Simeonov, and Chris
  Taggart.
\newblock The {euBusinessGraph} ontology: A lightweight ontology for
  harmonizing basic company information.
\newblock \emph{Semantic Web}, 13\penalty0 (1):\penalty0 41--68, 2022.
\newblock \doi{10.3233/sw-210424}.
\newblock Crossref registers the issued date as 2021 (IOS Press pre-press); the
  issue of record is volume 13 (2022). Ontology file inspected directly at
  \url{https://github.com/euBusinessGraph/eubg-data}: 34 object properties,
  none expressing ownership, parenthood or corporate-group membership.

\bibitem[Smart and Dudas(2007)]{smart2007purchasingsynergy}
Alan Smart and Andreas Dudas.
\newblock Developing a decision-making framework for implementing purchasing
  synergy: a case study.
\newblock \emph{International Journal of Physical Distribution \& Logistics
  Management}, 37\penalty0 (1):\penalty0 64--89, 2007.
\newblock \doi{10.1108/09600030710723327}.

\bibitem[Soylu et~al.(2022)Soylu, Corcho, Elves{\ae}ter, Badenes-Olmedo,
  Blount, Yedro~Mart{\'\i}nez, Kovacic, Posinkovic, Makgill, Taggart, Simperl,
  Lech, and Roman]{soylu2022theybuyforyou}
Ahmet Soylu, Oscar Corcho, Brian Elves{\ae}ter, Carlos Badenes-Olmedo, Tom
  Blount, Francisco Yedro~Mart{\'\i}nez, Matej Kovacic, Matej Posinkovic, Ian
  Makgill, Chris Taggart, Elena Simperl, Till~C. Lech, and Dumitru Roman.
\newblock {TheyBuyForYou} platform and knowledge graph: Expanding horizons in
  public procurement with open linked data.
\newblock \emph{Semantic Web}, 13\penalty0 (2):\penalty0 265--291, 2022.
\newblock \doi{10.3233/sw-210442}.
\newblock Full text inspected: supplier reconciliation to OpenCorporates via
  owl:sameAs; zero occurrences of ``subsidiary'', ``ultimate'' or
  ``ownership''; no gold standard and no reconciliation accuracy reported.

\bibitem[Steiner et~al.(2026)Steiner, Peeters, and Bizer]{steiner2026madibench}
Aaron Steiner, Ralph Peeters, and Christian Bizer.
\newblock {MaDI-Bench}: An end-to-end data integration benchmark.
\newblock \emph{arXiv preprint arXiv:2606.30371}, 2026.
\newblock URL \url{https://arxiv.org/abs/2606.30371}.
\newblock Verified at metadata level only (DBLP CoRR record); contents not
  read.

\bibitem[Thirumuruganathan et~al.(2021)Thirumuruganathan, Li, Tang, Ouzzani,
  Govind, Paulsen, Fung, and Doan]{thirumuruganathan2021blocking}
Saravanan Thirumuruganathan, Han Li, Nan Tang, Mourad Ouzzani, Yash Govind,
  Derek Paulsen, Glenn Fung, and AnHai Doan.
\newblock Deep learning for blocking in entity matching: A design space
  exploration.
\newblock \emph{Proceedings of the VLDB Endowment}, 14\penalty0 (11):\penalty0
  2459--2472, 2021.
\newblock \doi{10.14778/3476249.3476294}.

\bibitem[Thoma et~al.(2010)Thoma, Torrisi, Gambardella, Guellec, Hall, and
  Harhoff]{thoma2010harmonizing}
Grid Thoma, Salvatore Torrisi, Alfonso Gambardella, Dominique Guellec, Bronwyn
  Hall, and Dietmar Harhoff.
\newblock Harmonizing and combining large datasets --- an application to
  firm-level patent and accounting data.
\newblock NBER Working Paper 15851, National Bureau of Economic Research, 2010.

\bibitem[Tu et~al.(2023)Tu, Fan, Tang, Wang, Li, Du, Jia, and
  Gao]{tu2023unicorn}
Jianhong Tu, Ju~Fan, Nan Tang, Peng Wang, Guoliang Li, Xiaoyong Du, Xiaofeng
  Jia, and Song Gao.
\newblock Unicorn: A unified multi-tasking model for supporting matching tasks
  in data integration.
\newblock \emph{Proceedings of the ACM on Management of Data}, 1\penalty0
  (1):\penalty0 1--26, 2023.
\newblock \doi{10.1145/3588938}.

\bibitem[{U.S. Department of the Treasury, Bureau of the Fiscal
  Service}(2026)]{usaspending_api}
{U.S. Department of the Treasury, Bureau of the Fiscal Service}.
\newblock {USAspending} {API} v2 --- recipient endpoints.
\newblock \url{https://api.usaspending.gov/}, 2026.
\newblock API contract at
  \url{https://github.com/fedspendingtransparency/usaspending-api}. Accessed
  2026-08-27. recipient\_level is documented as R (neither parent nor child), P
  (parent recipient), C (child recipient); parent\_uei, parent\_duns,
  parent\_name and parent\_id are nullable; parents is an array; the structure
  is one hop deep with no grandparent chain.

\bibitem[Vitali et~al.(2011)Vitali, Glattfelder, and
  Battiston]{vitali2011network}
Stefania Vitali, James~B. Glattfelder, and Stefano Battiston.
\newblock The network of global corporate control.
\newblock \emph{PLoS ONE}, 6\penalty0 (10):\penalty0 e25995, 2011.
\newblock \doi{10.1371/journal.pone.0025995}.

\bibitem[Wang et~al.(2021)Wang, Li, and Hirota]{wang2021machamp}
Jin Wang, Yuliang Li, and Wataru Hirota.
\newblock Machamp: A generalized entity matching benchmark.
\newblock \emph{arXiv preprint arXiv:2106.08455}, 2021.
\newblock URL \url{https://arxiv.org/abs/2106.08455}.

\bibitem[Wei et~al.(2024)Wei, Karina, Chung, Jiao, Papay, Glaese, Schulman, and
  Fedus]{wei2024simpleqa}
Jason Wei, Nguyen Karina, Hyung~Won Chung, Yunxin~Joy Jiao, Spencer Papay,
  Amelia Glaese, John Schulman, and William Fedus.
\newblock Measuring short-form factuality in large language models.
\newblock \emph{arXiv preprint arXiv:2411.04368}, 2024.
\newblock URL \url{https://arxiv.org/abs/2411.04368}.
\newblock Introduces SimpleQA. Grading scheme: correct / incorrect / not
  attempted.

\bibitem[Wei et~al.(2025)Wei, Sun, Papay, McKinney, Han, Fulford, Chung,
  Tachard~Passos, Fedus, and Glaese]{wei2025browsecomp}
Jason Wei, Zhiqing Sun, Spencer Papay, Scott McKinney, Jeffrey Han, Isa
  Fulford, Hyung~Won Chung, Alex Tachard~Passos, William Fedus, and Amelia
  Glaese.
\newblock {BrowseComp}: A simple yet challenging benchmark for browsing agents.
\newblock \emph{arXiv preprint arXiv:2504.12516}, 2025.
\newblock URL \url{https://arxiv.org/abs/2504.12516}.

\bibitem[Zeakis et~al.(2023)Zeakis, Papadakis, Skoutas, and
  Koubarakis]{zeakis2023pretrained}
Alexandros Zeakis, George Papadakis, Dimitrios Skoutas, and Manolis Koubarakis.
\newblock Pre-trained embeddings for entity resolution: An experimental
  analysis.
\newblock \emph{Proceedings of the VLDB Endowment}, 16\penalty0 (9):\penalty0
  2225--2238, 2023.
\newblock \doi{10.14778/3598581.3598594}.

\bibitem[Zhang et~al.(2024)Zhang, Dong, Xiao, and Oyamada]{zhang2024jellyfish}
Haochen Zhang, Yuyang Dong, Chuan Xiao, and Masafumi Oyamada.
\newblock Jellyfish: Instruction-tuning local large language models for data
  preprocessing.
\newblock In \emph{Proceedings of the 2024 Conference on Empirical Methods in
  Natural Language Processing (EMNLP)}, pages 8754--8782. Association for
  Computational Linguistics, 2024.
\newblock \doi{10.18653/v1/2024.emnlp-main.497}.

\bibitem[Ziv et~al.(2022)Ziv, Gronau, and Fire]{ziv2022companyname2vec}
Ran Ziv, Ilan Gronau, and Michael Fire.
\newblock {CompanyName2Vec}: Company entity matching based on job ads.
\newblock In \emph{2022 IEEE 9th International Conference on Data Science and
  Advanced Analytics (DSAA)}, pages 1--10. IEEE, 2022.
\newblock \doi{10.1109/dsaa54385.2022.10032350}.
\newblock AUTHOR-NAME DISCREPANCY: arXiv and the IEEE DOI record give ``Ran
  Ziv''; OpenAlex renders it ``Ziv Ran''. arXiv/IEEE form used here.

\end{thebibliography}
\end{document}